\documentclass[fleqn,twoside]{article}
\usepackage{espcrc2}

\usepackage{graphicx}

\newcommand{\AmS}{{\protect\the\textfont2
  A\kern-.1667em\lower.5ex\hbox{M}\kern-.125emS}}

\title{Lateral distribution function of high energy muons in EAS around the knee}

\author{V. B. Petkov\address{Institute for Nuclear Research of RAS, Baksan Neutrino Observatory, 361609, Neutrino, KBR, Russia}%
        \thanks{vpetkov@yandex.ru},
        I. Alikhanov\addressmark,
        J. Szabelski\address{The Andrzej Soltan Institute for Nuclear Studies, Cosmic Ray Laboratory, IPJ, 90-950 Lodz 1, Box
447, Poland}
        }

\begin{document}

\begin{abstract}
The lateral distribution function of high energy  muons in EAS around the knee ($5.9
\le \lg N_e \le 7.1$) has been measured for near vertical showers ($\theta \le
20^{\circ}$, effective muon threshold energy is 230 GeV). The measurements have been performed
at the Baksan Underground Scintillation Telescope (BUST).
The electromagnetic component is measured by the "Andyrchy"
EAS array, located above the BUST. The knee in EAS size spectrum is found to be at $\lg
N_e \approx 6.3$. The experimental results are compared with Monte Carlo simulations.
\vspace{1pc}
\end{abstract}

\maketitle
\section{Introduction}
Measuring the lateral distribution function of high energy muons and its dependence on the shower size
provides additional information for studying the mass composition of the primary cosmic radiation as well
as for choosing a model of high energy hadron interactions. In this work we present results of measurements
of the muon lateral distribution functions in EAS for the following three regions around the knee: $5.9 \le
\lg N_e <6.3$, $6.3 \le \lg N_e < 6.5$ and $6.5 \le \lg N_e \le 7.1$ (the muon energy threshold is 230
GeV). The measurements have been performed using the facilities of the Baksan Neutrino Observatory: the
BUST and the "Andyrchy" EAS array~\cite{Alekseev93,Alekseev79}. The live data taking time is 24460.8 hours.
The knee position in the EAS spectrum corresponding to the considered zenith angles $\theta \le 10^{\circ}$
is found to be at $\lg N_e \approx 6.3$. We have also simulated the development of EAS in the Earth
atmosphere by means of the CORSIKA package (version 6720). As the high and low energy hadronic interaction
models we used QGSJET 01C and GHEISHA 2002d, respectively \cite{Heck98}. The simulations have been carried
out for primary protons and iron nuclei and the obtained muon lateral distribution function is compared
with the experimental one.

\section{Experiment}
The "Andyrchy" EAS array is located on the slope of the mountain Andyrchy above the BUST and consists of 37
plastic scintillation detectors of 1 m$^2$ area each. The distance between the detectors is about 40 m in
projection to the horizontal plane and the overall area of the installation is $4.5\cdot10^4$ m$^2$. The
detectors are arranged in such a way that the central one is just above the BUST at a vertical distance of
about 360 m which corresponds to 2060 m above the sea level. The shower trigger condition requires four
detectors to fire within 3 microseconds. The trigger's rate is about 9 s$^{-1}$. The array and its
characteristics are described in more details in \cite{Petkov2006}.

The BUST \cite{Alekseev79} is a four-floor building with $16.1 \times 16.1 \times 11.2$ dimensions located
in a mine at the effective depth 850 hg/cm$^2$. The floors as well as the four vertical sides of the
building are fully covered with 3150 liquid scintillation detectors. Each of  the detectors has the
dimensions ($0.7 \times 0.7 \times 0.3$) m$^3$. The telescope allows to determine the number of the passing
muons (1 - 200), their coordinates (with 0.7 m accuracy) and the arrival direction (with $1.5$ degree
accuracy). The coincidence trigger rate of the BUST and "Andyrchy" is about 0.1 s$^{-1}$.

The size, axis position and the EAS arrival direction are determined using the "Andyrchy" array data;
the BUST data are used to determine the number of muons crossing it. In the present analysis, only the
showers with axes in central part of the installation (the distance from the center is not larger than
50 m) and with $5.9\le\lg N_e\le7.1$ were taken into account. The accuracy of the determination of
$N_e$ for such showers is not worse than 12\%. For more details on identification of EAS parameters at
the "Andyrchy" detector array see, for example, \cite{Petkov2002,Petkov2003,Voevodsky97}.

\section{The fraction of muons in the telescope}
The underground telescope measures only a part of the total number of muons in EAS and the uncertainty
in the determination of the EAS axis position at the observation level is comparable with the size of
the BUST. In our experiment the fraction of muons in the telescope $\Delta(R)$ was measured as a
function of the distance $R$ between the center of the telescope and the EAS axis for a set of showers
with a given value of $N_e$ \cite{Chudakov97,Chudakov2000}. This fraction for a given range of $N_e$
is defined by

\begin{equation}
\Delta(R) = \frac{\overline n(R)}{\overline N_{\mu}(N_e)},
\end{equation}

 \noindent where $\overline n(R)$ is the mean number of muons in the telescope at the distance $R$, $\overline
N_{\mu}(N_e)$ is the mean number of muons in EAS. The mean number of muons in the BUST is determined as
follows. Events within a given range of $N_e$ are grouped according to the distance to BUST center with
step $\delta R = 10$ m. For each group, the number of muons is

\begin{equation}
M(R_i) = \sum_{j=1}^{K_i}{m_{ij}}
\end{equation}

\noindent where  $K_i = K(R_i)$ is the number of EAS in the $i$-th group,
 and $m_{ij}$ is the total number of muons in the BUST in the $i$-th group for the $j$-th EAS.

Thus, the mean number of muons for a given range of $N_e$ can be written as

\begin{equation}
{\overline n}(R_i) = \frac{M(R_i)}{K_i} = \frac{1}{K_i}\sum_{j=1}^{K_i}{m_{ij}}.
\end{equation}

 \noindent The mean number of muons in a shower is then calculated  as

\begin{equation}
{\overline N}_{\mu} = \frac{1}{S_t}\cdot \sum_i {\overline n}(R_i)\cdot S_r(R_i),
\end{equation}

\noindent where $S_t = 200$ m$^2$ is the effective area of the telescope, $S_r(R_i)$ is the area of the
ring of radius $R_i$ with $\delta R = 10$ m.

\begin{figure}[tb]
\vspace{9pt}
\resizebox{0.5\textwidth}{!}{%
\includegraphics{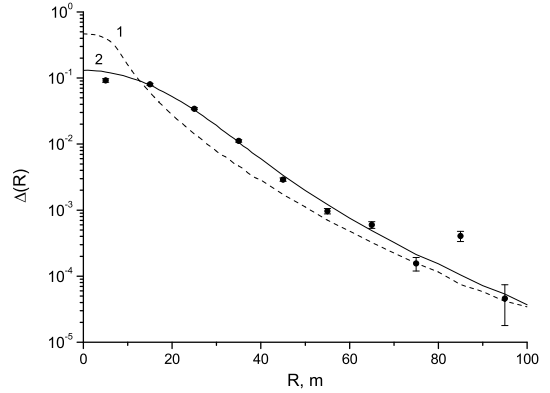}
} \caption{The mean number of muons in the telescope as a function of the distance $R$ between
 the center of the telescope and the shower axis
for showers with $5.9 \le \lg N_e <6.3$. The experimental measurements are depicted by the dots.
 The curves correspond to the calculations with
$r_0=4.6$ m using equation~\ref{eq:fmulateral} and for
  $\sigma_R=0$ m (curve 1) and  $\sigma_R=12$ m (curve 2).} \label{fig1}
\end{figure}

We have mentioned above that the fraction of muons with $E_{\mu} \ge 230$ GeV in the telescope as a
function of $R$ was obtained for the following three energy ranges about the knee in the EAS spectrum. For
the first one, before the knee $5.9 \le \lg N_e <6.3$ ($\lg {\overline N_e} = 6.1$), the mean number of
muons in a shower is $\overline N_{\mu} = 100\pm 2$; for $6.3 \le \lg N_e < 6.5$ ($\lg {\overline N_e} =
6.4$), this is $\overline N_{\mu} = 184\pm 9$; and for $6.5 \le \lg N_e \le 7.1$ ($\lg {\overline N_e} =
6.7$)  $\overline N_{\mu} = 334\pm 19$. The experimentally measured dependence $\Delta(R)$ for $5.9 \le \lg
N_e <6.3$ is shown in Fig. \ref{fig1} as black dots.

\section{The muon lateral distribution function}

The fraction of muons in the telescope  $\Delta(R)$ depends only on the muon lateral distribution function
$f(r)$ and the geometry of the installation. For distances much larger than the typical size of the
installation ($R \gg \sqrt{S_t}$), the fraction of muons is $\Delta(R)\approx S_t\cdot f(R)$. In the
analysis of the experimental data, the following form of the function was used

\begin{equation}
f(r) \sim \left({\displaystyle \frac{r}{r_0}}\right)^ {-\left({\displaystyle a + b\cdot
\left(\frac{r}{r_0}\right)^c}\right)}, \label{eq:fmulateral}
\end{equation}

\noindent which, according to the BUST data on the muon groups, sufficiently reproduces the lateral
distribution of high energy muons with the threshold energy-dependent parameter $r_0$, when  $a=0.58$,
$b=0.64$, $c=0.47$ \cite{Bakatanov83,Voevodsky93}. The same function fits the lateral distribution of muons
with the threshold energy 230 GeV in the simulated showers as well. However, in order to keep the good
agreement at the primary energies above $10^5$ GeV per nucleon, one has to change the values of the
parameters to: $a=0.65$, $b=0.72$, $c=0.4$, while $r_0$ depends only on the primary energy per nucleon.

The calculation of $\Delta(R)$ has been performed by the Monte Carlo method for a few values of $r_0$ with
taking into account the uncertainty in determination of the shower axis which, at the BUST level, is
$\sigma_R \approx 12$ m (the main contribution is from the determination of the EAS arrival direction due
to the large distance between the installations).

\begin{figure}[htb]
\vspace{9pt}
\resizebox{0.5\textwidth}{!}{%
\includegraphics{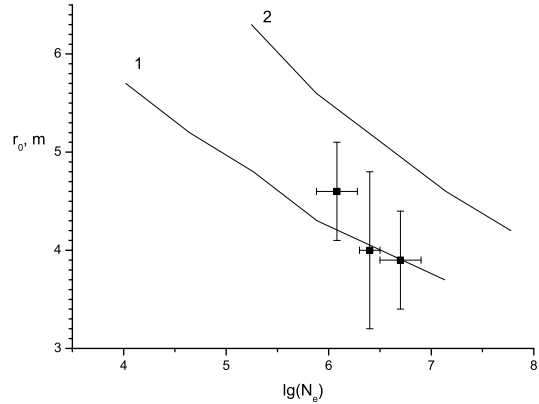}
} \caption{Dependence of the parameter $r_0$ in the muon lateral distribution function on the EAS size.
Points - experiment, lines - simulations for: 1 - primary protons, 2 - primary iron nuclei.} \label{fig2}
\end{figure}

In  Fig. \ref{fig1} we show the experimentally measured $\Delta(R)$ for $5.9 \le \lg N_e <6.3$ in
comparison with the calculations at $r_0 = 4.6$ m for two values of $\sigma_R$: 0 m (no uncertainty in the
determination of the arrival direction) and 12 m. This comparison gives the value of $r_0$ for each of the
ranges of $N_e$ (Fig. \ref{fig2}): for $\lg {\overline N_e} = 6.1$ $r_0 = 4.6\pm 0.5$ m, for  $\lg
{\overline N_e} = 6.4$ $r_0 = 4.0\pm 0.8$ m and for $\lg {\overline N_e} = 6.7$ $r_0 = 3.9\pm 0.5$ m.

The dependence of the experimental values of $r_0$ on  $\lg {\overline N_e}$ in comparison with the
corresponding calculations for protons and iron nuclei is displayed in Fig. \ref{fig2}. In this case the
mean values of $N_e$ are taken as the means for every fixed primary energy, without taking fluctuations and
the form of the primary spectrum into account.

\bigskip
This work was supported by the "Neutrino Physics" Program for Basic Research of the Presidium of the
Russian Academy of Sciences and by "State Program for Support of Leading Scientific Schools" (project no.
NSh-321.2008.2). This work was also supported in part by the Russian Foundation for Basic Research (grants
06-02-16135 and 08-07-90400).


\begin{thebibliography}{9}
\bibitem{Alekseev93}  E.N. Alexeyev et al., Izv. RAS, Ser. Phys. 57 (4) (1993) 99.

\bibitem{Alekseev79}  E.N. Alexeyev et al.,  Proc. 16th ICRC, Kyoto, 1979, v.10, p.276.

\bibitem{Heck98} D. Heck et al., Report FZKA 6019 (1998), Forschungszentrum, Karlsruhe.

\bibitem{Petkov2006} V.B. Petkov  et al., Instruments and Experimental Techniques, 2006, v. 49, No. 6, p. 785.

\bibitem{Petkov2002} V.B. Petkov  et al.,  Izv. RAS, Ser. Phys. 66 (11) (2002) 1560.

\bibitem{Petkov2003} V.B. Petkov et al.,  Proc. 28th ICRC,  Tsukuba, 2003, v.1, p.65.

\bibitem{Voevodsky97}  A.V. Voevodsky et al., Izv. RAS, Ser. Phys. 61 (3) (1997) 496.

\bibitem{Chudakov97} A.E. Chudakov et al.,  Proc.25 ICRC, Durban, 1997, v.6, p.173.

\bibitem{Chudakov2000} A.E. Chudakov et al.,  Proc. Xth International School "Particles and
Cosmology", Moscow, 2000, p.303.

\bibitem{Bakatanov83} V.N. Bakatanov et al.,  Proc. 18 ICRC, Bangalore, 1983, v.11, p.453.

\bibitem{Voevodsky93}  A.V. Voevodsky et al., Yad. Fiz. 56 (1993) 143.


\end{thebibliography}
\end{document}